\documentclass[journal]{IEEEtran}
\usepackage{xcolor,soul,framed} 
\colorlet{shadecolor}{yellow}
\usepackage[pdftex]{graphicx}
\graphicspath{{../pdf/}{../jpeg/}}
\DeclareGraphicsExtensions{.pdf,.jpeg,.png"}
\usepackage[cmex10]{amsmath}

\usepackage{array}
\usepackage{mdwmath}
\usepackage{mdwtab}
\usepackage{eqparbox}
\usepackage{graphicx}
\usepackage{url}
\usepackage{pifont}
\usepackage{times}
\usepackage{comment}

\begin{document}
\title{Considerations on the Design of Transceivers for Ambient Internet of Things}
\author{Yuxiao Zhao*,~\IEEEmembership{Graduate Student Member,~IEEE,}
      Zhen Shen, Shiyu Li, Jing Feng,~\IEEEmembership{Student Member,~IEEE}\\
      and Hao Min*,~\IEEEmembership{Member,~IEEE}

\thanks{Manuscript received January 10, 2024.}
\thanks{Yuxiao Zhao,  Zhen Shen, Shiyu Li, Jing Feng, and Hao Min are with Auto-ID Laboratory, Fudan University, Shanghai 201203, China (e-mail: zhaoyx19@fudan.edu.cn, hmin@fudan.edu.cn).}
} 

\markboth{IEEE XXX, JANUARY 2024
}{Yuxiao Zhao \MakeLowercase{\textit{et al.}}: Considerations on the Design of Transceivers for Ambient Internet of Things}

\maketitle

\begin{abstract}
The Ambient IoT (A-IoT) will introduce trillions of connections and enable low-cost battery-less devices. The A-IoT nodes can achieve low cost ($\sim\$ 0.1$ like RFID tag), sub-1mW average power consumption, $\leq 10$ kbps data rates, maintenance-free working for decades, cm-scale size, and support applications like supply chain and smart agriculture. The transceiver challenges in A-IoT focus on sub-mW receivers and crystal-less clock generation. The paper proposes an approximate low-IF receiver and carrier-auxiliary IF feedback LO synthesizer architecture for Type-B/C A-IoT devices, which tracks the RF carrier frequency and eliminates external crystals. The proposed receiver and LO generator are implemented using 55nm CMOS technology. After locking the LO calibration loop, the receiver sensitivity is better than -88 dBm. The proposed receiver architecture will promote zero-power devices for ubiquitous IoT connectivity, bridging digital and physical worlds.
\end{abstract}

\IEEEpeerreviewmaketitle

\section{Introduction}

\IEEEPARstart{I}{n} recent years, the number of Internet of Things (IoT) connections has exploded exponentially. According to IoT Analytics, by 2030, there will likely be more than 41 billion IoT connections \cite{2024iotanalytics}. In the future, trillions of nodes will enable massive IoT \cite{2021jiot}, which has already been discussed in massive machine-type communications (mMTC) of the fifth generation (5G) mobile communication \cite{2018_3GPP_Release14_MTC_overview, 2024_5G_Passive_IoT}. Traditional power supply solutions, such as wire-line power or batteries, are impractical for trillions of connections \cite{2024ambient_Access}. Meanwhile, a new IoT concept, Ambient IoT(A-IoT), is becoming popular \cite{2024_5G_Passive_IoT,2024ambient_Access,2024ambient_Agriculture,2024ambient_ECNCT,2024ambient_ICC_Workshop,2024ambient_MJIoT,2025_3gpp_Network_AmbIoT,2025_challenges_AmbIoT}.

The A-IoT achieves battery-free operation by introducing ambient energy harvesters to obtain energy from ambient sources, such as radio frequency, solar, thermal, vibration, pressure, etc. \cite{2024ambient_MJIoT}. The A-IoT devices are also called “zero power” devices due to their capability to operate without a dedicated power source \cite{2024zero-energy_device}. On the other hand, ambient communications utilizing backscatter modulation technology can establish long-term, maintenance-free, and energy-efficient networks due to the ultra-low power consumption characteristics \cite{2024ambient_Access}. The 3rd generation partnership project (3GPP) conducted a study in Release 18 to investigate ambient IoT use cases, deployment scenarios, and design targets, which will lead the Cellular A-IoT \cite{2023_3GPP_RAN, 2024_3GPP_RAN}.

Considering the large proportion of the A-IoT nodes in the entire network, this paper summarizes the design targets of these nodes and discusses the network characteristics accordingly.
\begin{itemize}
\item \textbf{Low Cost}: For IoT devices, the cost of radio frequency identification (RFID) tags is reaching $\$0.1$ after decades of optimization \cite{2022chiplessRFID_1centTag}. In contrast, the cost of typical IoT devices such as Cellular, Bluetooth, and WiFi ranges from $\$10$ to $\$100$. Therefore, the A-IoT aims to inherit the cost level of RFID Tags, making the scale of deployment close to the shipment volume of RFID, for example, 115 billion annually \cite{2024RFIDJournal}. In addition, the low cost means that the nodes have low complexity and few off-chip components such as crystals and batteries, which helps to keep a small volume factor and long running time to avoid maintenance \cite{2022griffithMSSC}.

\item \textbf{Low Power}: IoT technology has always been considered to be effective in promoting supply chain tracking and optimization, helping to achieve global carbon neutrality plans.
The A-IoT is energy-efficient with ultra-low-power consumption (Sub-1mW) radio operation, which supports sub-10 $\mu W$ average power consumption for decades of working life \cite{2024ambient_MJIoT, 2022griffithMSSC, 2025bridge_to_6G_ICM}. Based on the power consumption perspective, A-IoT will be the lowest energy consumption (per node) IoT.

\item \textbf{Low Data Rates}: In the 3GPP IoT space \cite{2024ambient_MJIoT}, the A-IoT node is described as the very low-end IoT with a peak rate not exceeding 10 kbps \cite{2024_5G_Passive_IoT}. The energy supply limits the data rate. And in the use case of A-IoT, such as supply chain management, precision agriculture, smart factories, etc., the common requirement is long-term maintenance-free, which gives up the communication speed \cite{2023_3GPP_RAN, 2024ambient_Agriculture}. In addition, the A-IoT nodes do not need to maintain continuous communication and will enable wake-up operation and duty-cycle radio. In terms of size, the A-IoT is a supplement to the global IoT network and can be called the cornerstone of IoT because of the massive connections. The A-IoT will connect the digital world and the physical world, achieving the ultimate goal of the IoT, ``Everything Connected'' \cite{2023uscoca}.

\item \textbf{Small Volume Factor}: To achieve high-density deployment, the A-IoT nodes need to have a small volume close to cm-scale, which will support large-scale sensing applications, such as underground sensing in precision agriculture\cite{2024ambient_Agriculture}.

\end{itemize}

3GPP discusses about 30 agreed use cases for the A-IoT application. This paper summarizes the following topics and makes key comparisons to better describe A-IoT's development trend.
\begin{itemize} 
\item[1)] \textbf{Sensing}: The A-IoT devices with sensors will construct a battery-free wireless sensor network \cite{2023MSNASSCC}, which can enable soil moisture monitoring, food, and vaccine quality control, structural health monitoring, smart house, etc. In this case, A-IoT nodes will primarily operate in transmission mode, like intermittent-working energy-limited beacons.

\item[2)] \textbf{Location}: Low-cost, high-precision indoor ($\sim 1m$) and outdoor ($\sim 10m$) positioning based on the A-IoT nodes will support many new applications, such as mall navigation, indoor drone navigation, tracking of products, tracking of the elderly, children, and tracking of livestock, etc. \cite{2023ambient_ICCC}.

\item[3)] \textbf{Supply Chain Management}: Traditional RFID is designed for rapid inventory and snapshot management in warehousing. However, A-IoT can track the assets and build a real-time online supply chain management system, which will optimize logistics and management costs \cite{2024Wiliot}.

\item[4)] \textbf{Actuator}: With an actuator, the A-IoT devices can achieve smart switches \cite{2017TCASIswitch} that can control the equipment in the farmland in smart agriculture or update the status information of medical instruments in hospital instruments management \cite{2023ambient_ICCC}.

\end{itemize}

Meanwhile, the A-IoT devices have been grouped into 3-type devices by the 3GPP RAN workgroup depending on the device complexity and power consumption level \cite{2023_3GPP_RAN, 2023ambient_ICCC}. However, the classification method proposed by 3GPP is designed for cellular A-IoT. The paper considers more open A-IoT issues and proposes the classification types of devices based on the power supply and communication activity. The proposed classification considers comprehensive evaluations from multiple standard organizations such as Bluetooth\cite{2024Bluetooth_AmbIoT}, IEEE\cite{2024IEEE802_11_AmbIoT}, and 3GPP \cite{2023_3GPP_RAN,2024_3GPP_RAN}. In addition, this paper also summarizes some chip design examples that meet the following categories.

\begin{itemize} 
\item \textbf{Type-A}: Passive Device, similar to Device 1 defined by 3GPP \cite{2024_3GPP_RAN}. Regarding power supply, the Type-A A-IoT device is a pure battery-free device without any energy storage capability. The ultra-high frequency (UHF) RFID ISO18000-6C (EPC Gen2) tag is the classical Type-A device. However, the A-IoT node needs more functions based on the RFID tag, such as sensing with higher receiver sensitivity and longer communication distance. In the communication type, the Type-A device also executes passive communication, which is also called backscatter communication, without any independent signal generation/amplification. Therefore, the type-A device has similar complexity compared with an EPC Gen2 tag, and some designs introduce a new radio frequency RF energy harvester to enhance tag communication distance for 5G applications\cite{2024JSSC_TypeA_Tag_5G_APP}. In summary, the Type-A device is similar to an RFID tag, which helps use mature RFID technology to design A-IoT. However, due to the backscatter communication, equivalent to OOK or ASK modulation, the type-A devices are not compatible with the popular IoT communication protocols (Bluetooth, WiFi, LoRa, NB-IoT, GSM, ...). The deployment of Type-A devices requires modifying and upgrading the existing base stations, which means more deployment costs.
\item \textbf{Type-B}: Semi-Passive Device, similar to Device 2a defined by 3GPP \cite{2024_3GPP_RAN}. Regarding communication activity, the type-B devices keep the backscatter transmitter, and the receiving path may be separated from the RF energy harvester. In addition, the Type-B has no independent signal generation but is potentially backscattering with reflection gain. Compared with Type-A, the Type-B A-IoT devices have limited energy storage capability and do not need to be replaced or recharged manually. Type-B devices can have a small battery with limited capacity, but the power supply must last several decades. Chip design research for Type-B devices has become popular in recent years. Some works introduce multiple antennas to design independent impedance matching for backscatter TX, receiver, and energy harvester \cite{2023ISSCC_Type-B,2024ISSCC_Type-B}. Many works have achieved good compatibility between backscatter communication and various communication protocols by the backscatter modulator, such as WiFi \cite{2021VLSIC_Type-B,2020ISSCC_Type-B,2023ISSCC_Type-B,2024ESSERC_Type-B,2022wifible_ISSCC_Type-B,2024wifi_JSSC_Type-B}, Bluetooth\cite{2023ESSCIRC_Type-B,2024ISSCC_Type-B,2024ESSERC_Type-B,2022wifible_ISSCC_Type-B,2023ISSCC_Type-B_ZhaoBo}, Zigbee \cite{2024ESSERC_Type-B}, Z-Wave \cite{2024ESSERC_Type-B}.
Although the reflector amplifier is proposed to extend the backscatter communication distance by introducing reflection gain, some work only reports on-board designs \cite{2014TMTT_reflection_amp, 2023LMWT_reflection_amp}. Currently, there is no silicon implementation. In addition, to improve the downlink communication distance and avoid the poor demodulation sensitivity brought by envelope detector and 1-bit quantizer in traditional RFID tags, the Type-B devices can introduce low power wake-up receiver technology\cite{2020ISSCC_Type-B,2023ISSCC_Type-B,2021VLSIC_Type-B,2024ISSCC_Type-B,2023ESSCIRC_Type-B,2023ISSCC_Type-B_ZhaoBo}. In summary, Type-B device is designed to achieve compatibility between backscatter communication and popular short-range IoT protocols, and achieve ambient energy harvesting and energy storage with high-efficiency power management circuit.

\item \textbf{Type-C}: Active Deivce, similar to Device 2c defined by 3GPP \cite{2024_3GPP_RAN}. The Type-C tags are active radio devices without backscatter communication compared to Type-A and Type-B. The Type-C devices have ambient energy harvesters and an optional small battery with a limited capacity. By introducing low-power radio technology and low-complexity circuit design, the Type-C device reduces over-design and only meets the loose wireless specification, which will reduce the cost of existing popular wireless IoT nodes\cite{2024TMTT_Type-C,2024Wiliot_Type-C}. In addition, Type-C devices' active radio power consumption during transmitting or receiving is less than 1 mW, which means less than 10 $\mu W$ average power consumption at a common $1\%$ duty cycle. The Type-C devices meeting popular communication protocols can achieve node-node communication, which is rarely reported in Type-B devices\cite{2023ISSCC_Type-B_ZhaoBo}.

\end{itemize}

Based on the above discussion, this paper focuses on the design of transceivers in A-IoT, including the development of the physical layer, radio frequency specifications, and feasible transceiver architectures. The paper proposes a crystal-less transceiver architecture for Type-B or Type-C devices. The proposed transceiver discusses an "approximate low-IF" receiver architecture and a "carrier-auxiliary IF feedback" LO frequency synthesizer for A-IoT applications.

The rest of this paper is organized as follows. In Section II, the paper introduces the physical layer protocol in A-IoT, including the downlink/uplink physical channel in 3GPP and the radio frequency specification of A-IoT. Section III describes the feasible transceiver architectures and some critical transceiver design considerations. Section IV introduces the proposed crystal-less transceiver architecture for Type-B or Type-C devices. Section V provides the detailed circuit designs of the ``approximate low-IF'' receiver architecture and ``carrier-auxiliary IF feedback'' LO frequency synthesizer. In Section VI, this paper introduces the key simulation and measurement results of the proposed transceiver and core circuits. Sections VII and VIII further discuss and conclude this paper.


\section{Evolutions of Physical Layer Protocol in Ambient IoT}

In the new A-IoT paradigm, defining physical layer protocols is crucial for low-cost node design, including communication carrier frequency, modulation, data rate, encoding formats, etc. In the downlink, the power consumption budget of the A-IoT tag receiver is limited to 1 mW, which is insufficient to support complex modulation, such as PSK and QAM, especially for carrier frequencies above 1 GHz\cite{2023OJSSCS_ULPRX}. For the uplink, it's possible to achieve low-power high-order modulation based on backscatter communication in the A-IoT device\cite{2020ISSCC_Type-B,2021VLSIC_Type-B,2022wifible_ISSCC_Type-B,2023ESSCIRC_Type-B,2024wifi_JSSC_Type-B,2023ISSCC_Type-B,2023ISSCC_Type-B_ZhaoBo,2024ESSERC_Type-B}. The following paragraphs will provide a summary description of A-IoT physical layer designs based on the 3GPP works.

\subsection {Downlink: Reader to Device (R2D) Communication}
A-IoT's physical layer design goal is to build a simple, low complexity, and low power consumption network. 3GPP defines one physical channel for the R2D link, called the Physical Reader to Device Channel (PRDCH). The channel transmits data and control information from the reader (base station) to the A-IoT device. The R2D link defines an OFDM-based OOK waveform with a subcarrier spacing of 15 kHz. The line codes are Manchester and PIE encoding. 3GPP also defines a carrier frequency offset calibration signal in the R2D link that can be used to synchronize/calibrate the device clocks, such as the LO frequency. In addition, the carrier frequency offset calibration signal will help to realize the ``carrier-auxiliary IF feedback'' LO frequency synthesizer proposed in this paper.

\subsection {Uplink: Device to Reader (D2R) Communication}
The D2R link also has one physical channel, the PDRCH, which carries data and control information. For D2R by backscattering, the external carrier wave provides the waveform. The D2R baseband modulations can be set to OOK, Binary PSK, Binary FSK, as MSK (and not GMSK), and use single-sideband or double-sideband modulation according to the application environment. The line codes are Manchester encoding, FM0 encoding, Miller encoding, and no line coding. For channel coding of D2R, convolutional codes are preferred.

\subsection {Radio Frequency Specification}
\begin{figure}[tp]
\centerline{ \includegraphics[width=3.6in]{"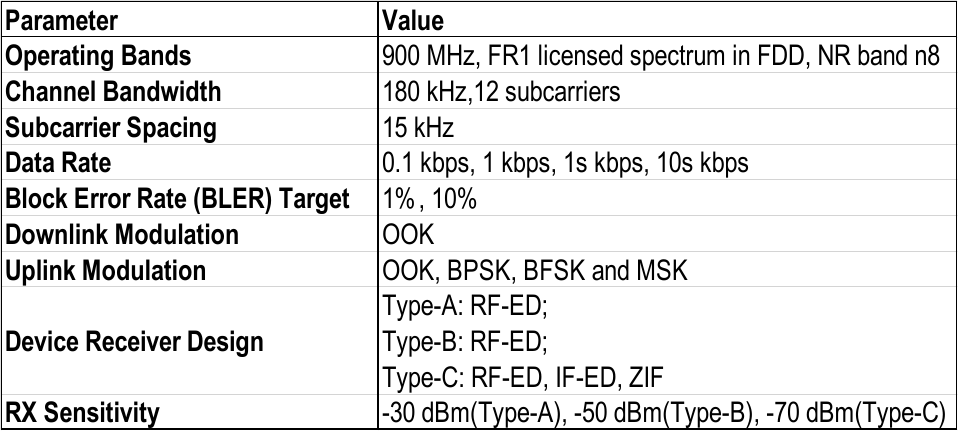"} }
\caption{Key RF Performance Parameters of Ambient IoT}
\vspace{-0.2in}
\label{fig1}
\end{figure}

The radio frequency specifications of A-IoT are summarized in Fig.~\ref{fig1} based on the 3GPP physical layer draft\cite{2024_3GPP_RAN} and some papers\cite{2025A-IoTPhysicalLayer_arXiV, 2025A-IoTPhysicalLayer_MC, 2025A-IoTPhysicalLayer_India}. According to the \cite{2024_3GPP_RAN_Revised_SID}, A-IoT uses FR1 licensed spectrum in FDD. The NR band n8 can be used as an example band. The channel bandwidth for the A-IoT system is 180 kHz, and the frequency spacing of the subcarrier is 15 kHz. The target data rate of the A-IoT prototype currently ranges from 0.1 to several tens of kbps\cite{2024_3GPP_RAN}. Therefore, the A-IoT is a narrowband Internet of Things. The bandwidth is close to NB-IoT, the data rate is close to LoRa, and the cost is close to RFID. Considering power consumption and performance, 3GPP has provided recommended receiver architecture designs for different device types, including RF-ED (RF envelope detector), IF-ED (IF envelope detector), and ZIF (Zero-IF) receivers. As a supplement, low-IF (LIF) and uncertain-IF receiver architectures are discussed in this paper. 3GPP has not yet given an exact value for the reference receiver sensitivity level. This paper proposes three recommendation levels of receiver sensitivity. The receiver sensitivity of Type-A devices is slightly better than Gen2 tags, reaching -30 dBm. The receiver sensitivity of Type-C devices is close to the minimum sensitivity of active radio with similar coverage of A-IoT, such as -70 dBm of Bluetooth. The receiver sensitivity of the Type-B is between the Type-A and Type-C, defined at -50 dBm.

\section{Low-power Transceiver Design Considerations for Ambient IoT}

In this section, the paper will introduce the low-power transceiver design considerations for A-IoT, including the recommended transceiver architectures for A-IoT in 3GPP, the design considerations for sub-mW receiver, and the key clock generation architecture.

\subsection {The Recommended Transceiver Architecture in 3GPP}
\begin{figure}[tp]
\centerline{ \includegraphics[width=3.6in]{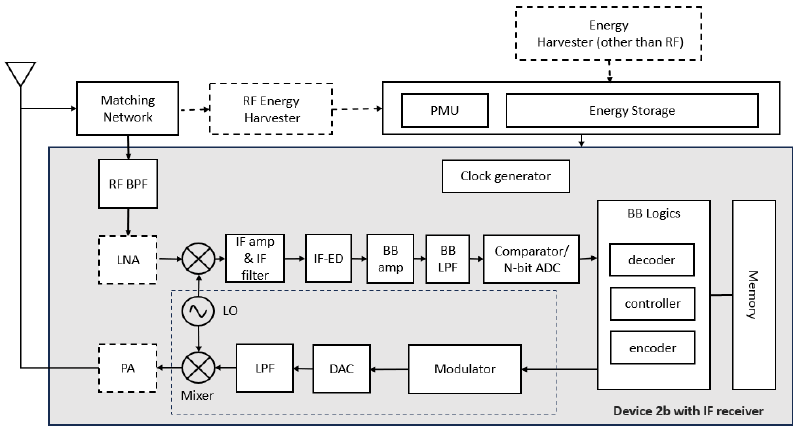} }
\caption{Zero-IF Receiver Architecture for A-IoT Type-C Devices}
\vspace{-0.2in}
\label{fig2}
\end{figure}

3GPP has defined various A-IoT device architectures for different device types \cite{2024_3GPP_RAN}. The Type-A (Device 1 in 3GPP) and Type-B (Device 2a in 3GPP) devices use an RF-ED receiver architecture. In addition, the Type-B may introduce LNA to improve the noise figure and enhance the receiving sensitivity. The most important feature of Type-B is the introduction of a reflector amplifier in the uplink backscatter path, which can significantly increase the communication distance between the device and the reader (base station). Type-C devices have various receiver architectures, including RF-ED, IF-ED, and ZIF. The IF-ED introduces a down-conversion stage before envelope detection, and the baseband also adds an N-bit ADC (N is a smaller integer), which helps to improve receiver sensitivity compared with the RF-ED receiver. 3GPP also recommends ZIF receiver architecture for Type-C devices (Device 2b in 3GPP), which is an energy-hungry implementation\cite{2021RFIC_Zero-IF_RX,2022JSSC_Zero-IF_SSPTRX}. The detailed Type-C device architecture based on the ZIF receiver is shown in Fig.~\ref{fig2}.

\subsection {The Design Considerations for Sub-mW Receiver}
Currently, there are many works on the transmitter design of the A-IoT devices\cite{2020ISSCC_Type-B,2021VLSIC_Type-B,2023ISSCC_Type-B,2024ISSCC_Type-B,2024ESSERC_Type-B,2023ESSCIRC_Type-B,2024wifi_JSSC_Type-B,2023ISSCC_Type-B_ZhaoBo,2022wifible_ISSCC_Type-B,2024TMTT_Type-C,2024Wiliot_Type-C}. Based on the above discussion, 3GPP has defined clear receiver architecture for Type-A and Type-B. This paper mainly discusses other possible receiver architectures for Type-C, which will supplement the implementation of A-IoT devices.

Two receiver architectures for low-power radio with higher sensitivity are mixer-first and LNA-first. Although the LNA-first receiver supports long-distance applications with -100 dBm receiver sensitivity, the LNA requires a mW power budget. In addition, the excessively high sensitivity design in A-IoT devices may mean over-design. The mixer-first receiver uses a mixer as the first stage, avoiding the active LNA. Although the mixer-first receiver suffers from a higher noise figure (NF), which reduces the receiver sensitivity, this architecture can still achieve considerable sensitivity and selectivity levels under sub-mW power budgets\cite{2020_ULPRX_CICC,2022_ULPRX_CICC,2023OJSSCS_ULPRX}.

In the sub-mW receiver designs, the selection of the IF is crucial for the receivers with a LO and a mixer. Since the OOK signal is its mirror, it does not cause demodulation issues. Therefore, the ZIF receiver does not require an I/Q mixer, which can save the power of the LO buffer\cite{2022_OJSSCS_ULPRX}. However, it is well known that ZIF receivers are affected by flickering noise. In addition, although the A-IoT standard designed by 3GPP uses an OOK modulation waveform, other standard organizations also consider the FSK modulation waveform \cite{2024Bluetooth_AmbIoT}. Uncertain-IF is a popular sub-0.1 mW receiver architecture, which uses a free-running oscillator and avoids the power consumption of the phase-locked loop. However, the wide-band IF path introduced to tolerate the poor frequency stability of the LO brings significant demodulation noise, resulting in a poor overall noise figure. The uncertain-IF architecture is invalid for IoT standards with multiple channel allocations due to the lack of PLL or FLL for channel switching \cite{2022_OJSSCS_ULPRX}. The low-IF receiver has always been a popular solution for low-power BLE receivers, but it has an image rejection issue \cite{2018TMTT_ULP_Low_IF_BLE,2015ISCAS_ULP_Low_IF_BLE}. The low-IF receiver must use a front-end image rejection filter or an I/Q mixer at the expense of higher LO buffer power \cite{2022_OJSSCS_ULPRX}.

\subsection {The Low-Cost Low-Power Clocks for A-IoT Transceiver}
According to the 3GPP definition, the clock requirements in A-IoT devices include five purposes \cite{2024_3GPP_RAN}. Clock purpose $\# 1$ is the sampling clock used for baseband signal processing. Clock purposes $\# 2$ and $\# 3$ are, respectively, small and large frequency offsets, which are used for backscatter modulation. Clock purpose $\# 4$ is a timing counting clock that controls device status. Clock purpose $\# 5$ is the local oscillator clock.

Based on the frequency ranges, these clocks can be divided into 10s kHz, 1s-10s MHz, and 100s-1000s MHz clocks. The 10s kHz clock is generally provided by an external low-frequency crystal oscillator, such as the 32 kHz real-time clock (RTC) crystal commonly used in IoT devices. 3GPP also points out that an on-chip calibrated relaxation oscillator can generate a 10s kHz clock with a 1000-10000 ppm frequency accuracy\cite{2024_3GPP_RAN}. In addition, the on-chip oscillator can also produce a 1s-10s MHz clock with relaxed precision requirements, such as 1000-10000 ppm.

The LO clock is in the 100s~1000s MHz frequency range with 10s~200ppm clock accuracy requirement in 3GPP\cite{2024_3GPP_RAN}. There are two popular LO generation solutions: FLL and PLL. The PLL achieves phase tracking and introduces high-power components such as TDC in DPLL and CP in CPPLL. The FLL is a simplified loop from PLL, operating in the frequency domain, and only achieves frequency tracking without phase noise suppression capability. In addition, there are two implementations for on-chip oscillators: ring and LC oscillators. LC oscillators have better phase noise performance with a large-area inductor. The area cost of the inductor is not significant for the SoC in type-C devices. However, the LC oscillators are not applicable in type-A devices because they cost similarly to the Gen2 tag. Some work uses wire-bonded or external inductors\cite{2023OJSSCS_ULPRX}, which are invalid in small-volume, low-cost A-IoT devices operating in extreme environments. Ring oscillators have poor phase noise, and frequency stability is greatly affected by PVT, but they have higher area efficiency. The phase noise of the LO has a significant impact on the sensitivity of the receiver. Still, it will not become a limiting factor for Type-C devices' -70 dBm sensitivity target.

The benefits of higher clock accuracy are obvious. Improved clock accuracy (with smaller uncertainty) allows smaller frequency guard bands, which improves spectrum efficiency and reduces power consumption. 3GPP also gives an example\cite{2024_3GPP_RAN}: Due to the influence of manufacturing processes and environments such as PVT, the initial carrier frequency offset (CFO) may reach 1000-10000 ppm, which will cause an offset of up to 900 kHz—9 MHz for the 900 MHz radio frequency carrier frequency. If the residual CFO after calibration can be reduced to 10s ppm, it means several tens of kHz guard bands in the 900 MHz band. Because the typical D2R transmission bandwidth is 10s or 100s kHz, the 10s kHz guard band setting is acceptable. In contrast, a CFO of 100s ppm requires a guard band of 100s kHz, which is unacceptable in the spectrum efficiency.

Low-cost clock solutions also need to consider the number of external components, such as the crystal. Designing A-IoT devices without any crystals has become a key issue, which has also been mentioned in 3GPP\cite{2024_3GPP_FL_Summary}. Currently, the crystal-less radio solutions all obtain the reference frequency from the RF wireless signal and calibrate the on-chip oscillators. \cite{2025ISCAS_XO_Less_YXZ} defines 4 types of crystal-less receiver architectures, and the Class-AB architecture reduces the number of high-power modules operating at radio frequency with the lowest power consumption. The receiver architecture proposed in this paper is improved based on the Class-AB crystal-less receiver.

\section{Proposed A-IoT Receiver Architecture}

\begin{figure}[tp]
\centerline{ \includegraphics[width=3.8in]{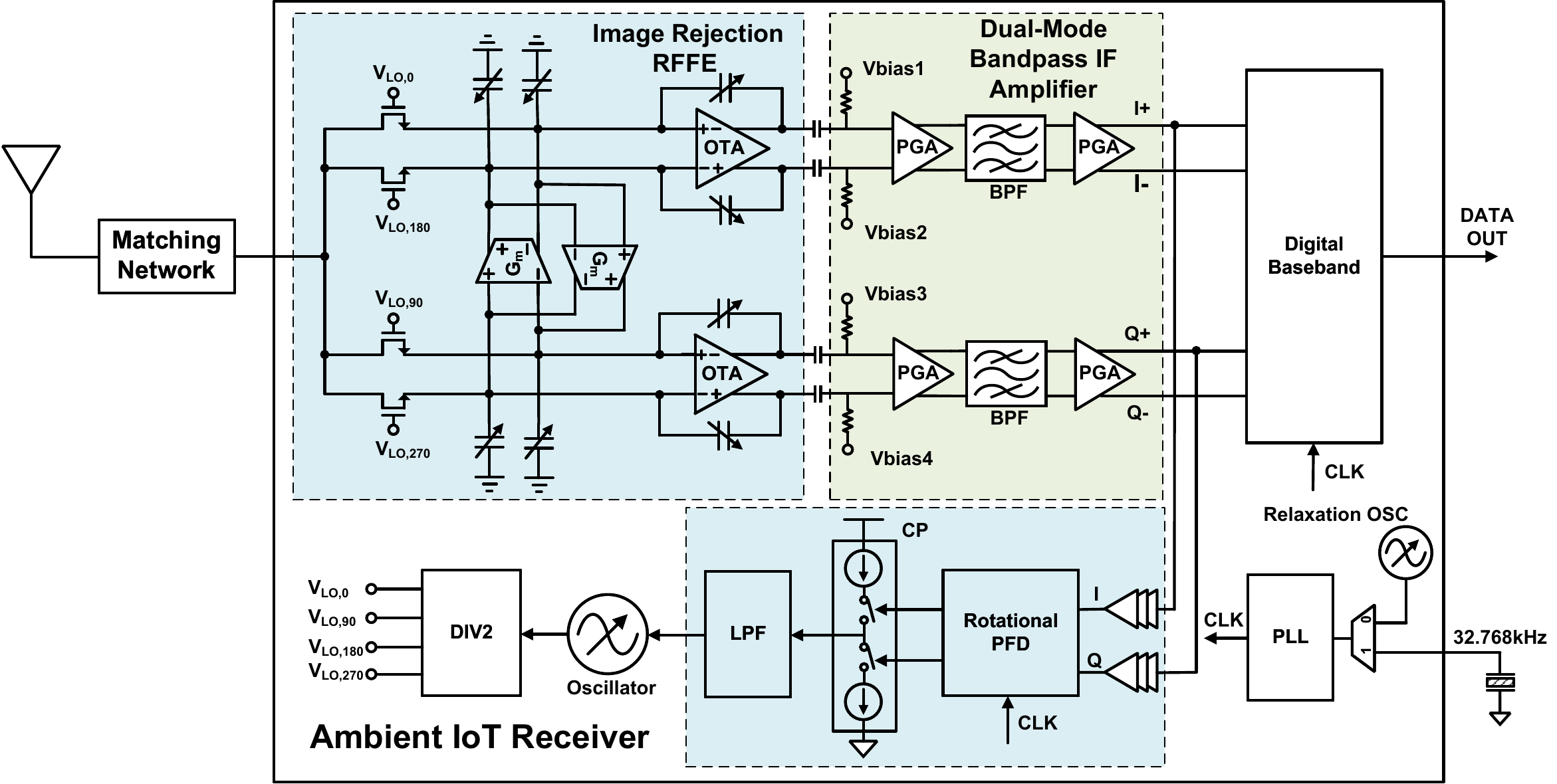} }
\caption{Proposed A-IoT Receiver Architecture for Type-B or Type-C Devices}
\vspace{-0.2in}
\label{fig3}
\end{figure}

\subsection {IF Negative Feedback Class-AB Crystal-less Receiver}

The proposed Class-AB crystal-less receiver architecture is shown in Fig.~\ref{fig3}. The frequency of IF is used as the negative feedback frequency signal in the frequency detector, $f_{FB}$. The reference frequency $f_{REF}$ is generated by a low-frequency frequency synthesizer, which uses a kHz on-chip relaxation oscillator with temperature compensation or an external 32kHz crystal oscillator (low-cost) as the original reference frequency, $f_{XO}$. Therefore, the LO frequency synthesizer is a cascade frequency synthesis system. The reported temperature-compensated kHz on-chip relaxation oscillator has a frequency stability of ±1000ppm (±0.1$\%$)\cite{2017_IFCS_RelaxOsc,2023_JSSC_RelaxOsc}. The output frequency range of the low-frequency synthesizer is 0.5-1.5MHz, which can provide an IF frequency with ±0.1$\%$ uncertainty (±1kHz for 1MHz IF). The RF front-end uses a 4-path passive mixer-first architecture to minimize power consumption as much as possible while providing interference rejection capability, which offers a good power and noise trade-off for low power receiver design\cite{2010TCASI_NPPM}. The proposed Class-AB crystal-less receiver has 3 working steps.

(1) Step A: Uncertain-IF Mode. The frequency drift/error of the oscillator is due to the combined effect of process and temperature. After temperature compensation based on a coarse frequency lookup table (LUT), the LO frequency drift can be below ±500 ppm (0.45MHz to 0.9GHz), which can be achieved in MEMS\cite{IUS2006_BAW} or LC oscillators\cite{UMichigan_alghaihab2020}. For the ring oscillators, the frequency drift may be higher than ±1000ppm (0.9MHz to 0.9GHz) after simple temperature compensation\cite{2010_Zhang_Xuan_PVT_Ring_Osc}. In the Step-A stage, due to the LO frequency uncertainty, the receiver works in the uncertain-IF mode, which has a wide IF bandwidth to adapt to LO frequency variation. The wider IF bandwidth will negatively affect the signal-to-noise ratio and reduce the receiver sensitivity.

(2) Step B: IF Feedback LO Calibration. After receiving the carrier with OOK modulation signal, the receiver uses a frequency detector to compare the $f_{FB}$ with the $f_{REF}$, controls the charge pump (CP) to calibrate the LO, and then reduces the frequency difference between the carrier and the LO. Finally, the carrier frequency interlocks with the IF and LO frequencies.

(3) Step C: Approximate Low-IF Mode. After obtaining the LO frequency with a smaller frequency deviation, the receiver will enable a low-bandwidth IF path to improve noise performance and sensitivity. Because of no phase noise suppression capability, the LO frequency calibration loop makes the oscillator work in the free-running mode. Compared with the low-IF receiver, the approximate low-IF receiver has similar RF performance and worse near-carrier LO phase noise.

\subsection {Carrier-Auxiliary IF Negative Feedback LO Frequency Calibration Loop}
\begin{figure}[tp]
\centerline{ \includegraphics[width=3.6in]{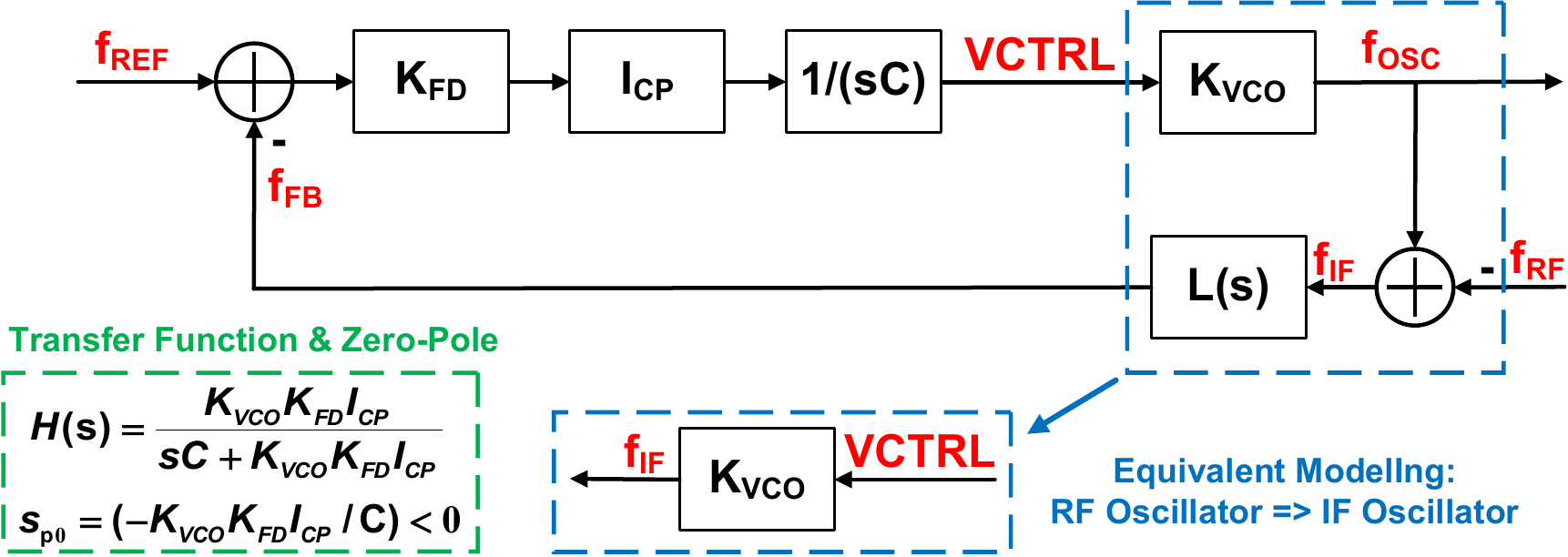} }
\caption{Simplified s-domain Model of LO Frequency Calibration Loop}
\vspace{-0.2in}
\label{fig4}
\end{figure}

The key technology in the crystal-less receiver is the LO frequency calibration loop as shown in Fig.~\ref{fig4}. The carrier-auxiliary IF negative feedback LO frequency calibration loop consists of RFFE (4-path passive mixer and trans-conductance amplifier), IF blocks (analog baseband), Schmitt trigger (square wave shaping), rotational frequency detector (RFD), charge pump, loop filter, and voltage-controlled oscillator (VCO). The RFD is a normal module in the clock data recovery application (RFD), achieves IF frequency detection, and can endure long-term ``0'' or ``1'' in the OOK waveform.

Fig.~\ref{fig4} shows the simplified s-domain model, closed-loop transfer function, and poles of the calibration loop in the frequency domain. The RF-VCO frequency is down-converted to the IF frequency, which can be equivalent to an IF VCO\cite{JSSC2022_RBFLL}. Although the loop locks the LO frequency and IF frequency to the accurate carrier frequency, the carrier plays only an auxiliary role in the loop dynamics. In other words, the carrier with OOK modulation helps the frequency detector to operate at MHz IF frequency rather than GHz LO frequency through down-conversion, avoiding higher power consumption in the LO feedback path. Therefore, this work introduces a carrier-auxiliary IF feedback calibration loop. In addition, there is no frequency conversion in the IF path, which means $L(s)=1$. The transfer function shows that the loop is a first-order loop in the frequency domain, which means that there are no loop stability issues after setting appropriate loop parameters.    

\subsection {Mixer-First RFFE with Image Rejection and Out-of-Band Interference Suppression}

This paper proposes a 4-path mixer-first RFFE with image rejection and out-of-band interference suppression. The passive mixer achieves baseband impedance mapping, which can construct a high-Q radio frequency band-pass filter at the LO frequency, realizing a SAW-less receiver and reducing the A-IoT devices' cost. The center frequency of the filter can be shifted based on the direction of the Gyrator, and the frequency variation can be tuned by the following formula:
\begin{equation}
\setlength{\abovedisplayskip}{0pt}
\setlength{\belowdisplayskip}{0pt}
\label{Gyrator}
\begin{split}
\Delta f=\frac{2G_m}{C}
\end{split}
\end{equation}

Therefore, the center frequency of the equivalent RF band-pass filter can be tuned by the size of trans-conductance (Gm) and capacitor (C), which will help achieve image signal rejection.

\subsection {The Target IF Planning}
Based on the above discussion, the LO frequency drift after LUT temperature compensation is ±500 ppm (0.45 MHz to 0.9 GHz). Because the channel bandwidth (CBW) of A-IoT is 180 kHz, the 450 kHz uncertainty will cover 2-3 channels. Therefore, the target IF frequency, $f_{IF}$, should meet:

\begin{equation}
\setlength{\abovedisplayskip}{0pt}
\setlength{\belowdisplayskip}{0pt}
\label{IF frequency1}
\begin{split}
f_{IF}>3 * 180kHz=540kHz
\end{split}\end{equation}

In addition, the image signal can be set in the available frequency guard band (3GPP is discussing) by finding an appropriate IF, which will help simplify image interference designs. Like the BLE receiver design in \cite{2014_Single_Path_BLE}, when only considering interference from other A-IoT devices, the IF can be set:
\begin{equation}
\setlength{\abovedisplayskip}{0pt}
\setlength{\belowdisplayskip}{0pt}
\label{IF frequency2}
\begin{aligned}
f_{IF}
&=\frac{CBW}{4}+\frac{CBW}{2}*n,n\in Z, n\geq 0\\
&=585, 675, 765, 855, 945, 1035kHz, ...
\end{aligned}
\end{equation}

Because too low IF frequency may introduce flick noise and DC offset issues like a zero-IF receiver, this paper chooses 1035 kHz as the target IF frequency.

\section{Circuits Design for the Proposed Receiver}

\subsection {4-path Passive Mixer-First RF Front-end}
\begin{figure}[tp]
\centerline{ \includegraphics[width=3.6in]{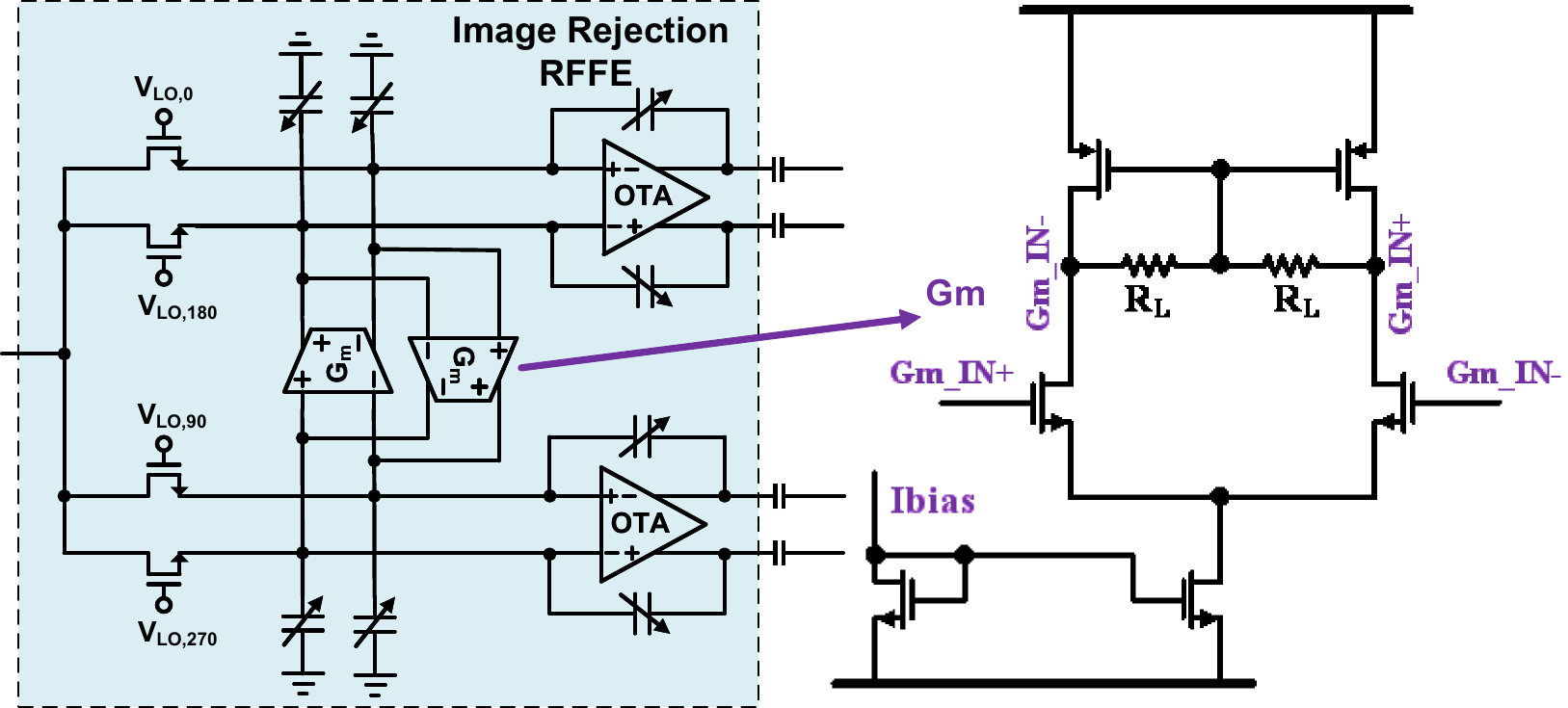} }
\caption{4-path Passive Mixer-First RF Front-end}
\label{fig5}
\end{figure}

The proposed RF front-end circuit is shown in Fig.~\ref{fig5}. The 4-path passive mixer and the Gyrator are combined to construct an RF band-pass filter with a tuning center frequency to achieve image rejection. The Gyrator is composed of trans-conductance amplifiers (Gm) and capacitors. The Gm adopts a common-source differential structure. By adjusting the bias current, the value of Gm can be changed, therefore, the adjustment of the center frequency can be realized. The TIA is the gain stage of the RFFE and uses the classical differential common-source structure. The common-mode feedback is achieved by the two resistors in the output node.

\subsection {Carrier-Auxiliary IF Feedback Crystal-Less LO Frequency Synthesizer}

\begin{figure}[tp]
\centerline{ \includegraphics[width=2in]{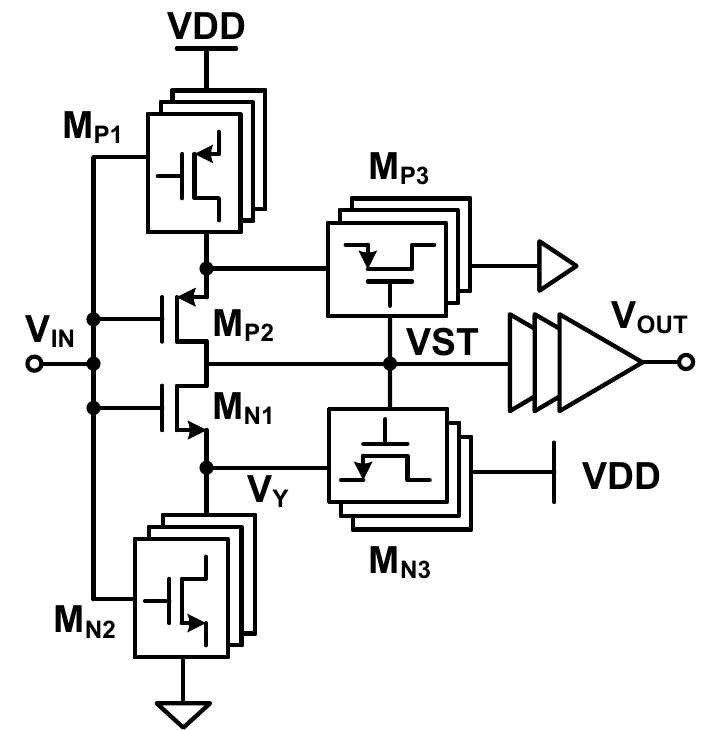} }
\caption{Schmitt Trigger with Programmable Threshold}
\label{fig6}
\end{figure}

\begin{figure}[tp]
\centerline{ \includegraphics[width=3.4in]{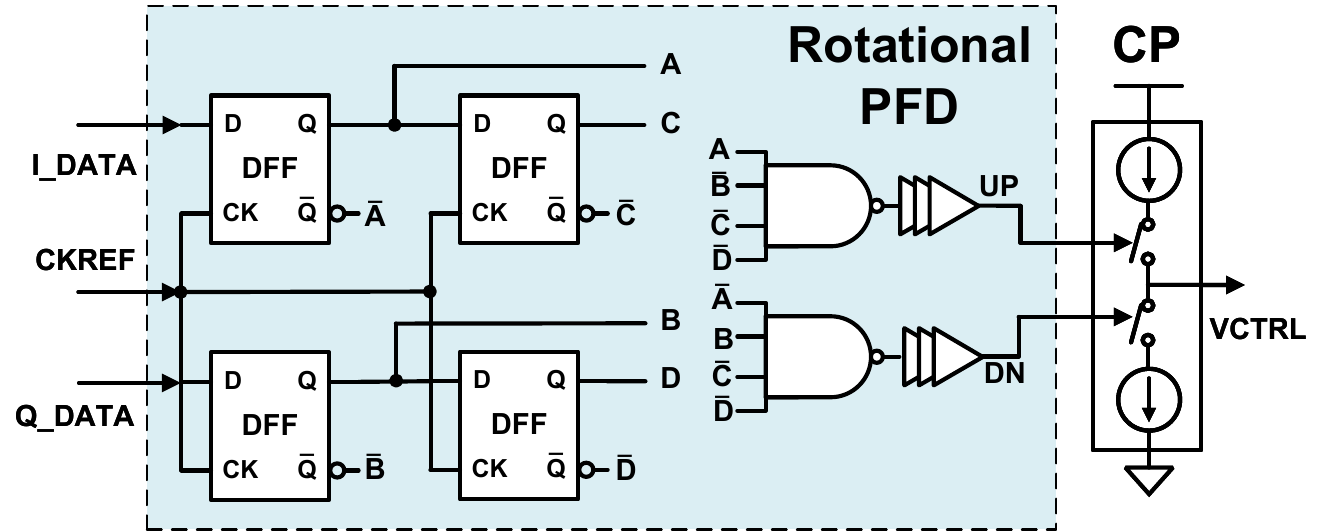} }
\caption{Rotating Frequency Detector}
\label{fig7}
\end{figure}

\begin{figure}[tp]
\centerline{ \includegraphics[width=2.8in]{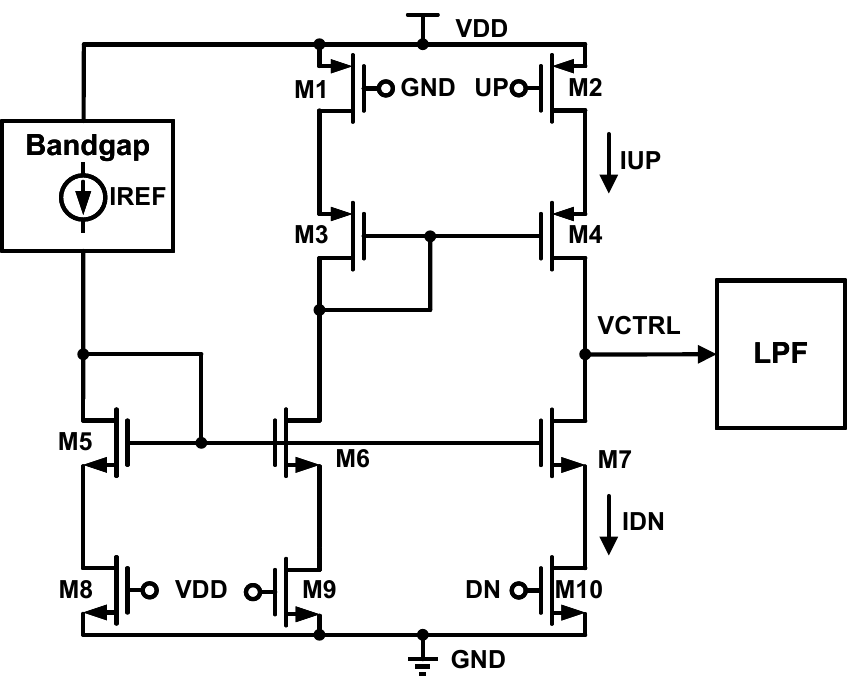} }
\caption{Source-Switch Charge Pump}
\label{fig8}
\end{figure}

The carrier-auxiliary IF feedback crystal-less LO frequency synthesizer consists of a Schmit trigger, a rotating frequency detector (RFD), a charge pump, a low-pass filter (LPF), and an oscillator. As shown in Fig.~\ref{fig6}, the designed Schmitt trigger achieves programmable threshold by tuning the digital-controlled MOSFET array and changing the equivalent size. The RFD in Fig.~\ref{fig7} uses the reference clock to sample the I/Q data signal. Then, two AND gates generate pulses based on the 4 edges. If the I/Q data frequency is lower than the reference clock, a down pulse is generated at the DN node and vice versa. The charge pump in Fig.~\ref{fig8} adopts a source-switch structure, which has a faster switching speed and has a smaller dynamic mismatch. The on-chip LO is generated from the 2-stage ring VCO, which consists of two differential trans-conductance units and a cross-coupled RC network\cite{2019_ESSCIRC_2_stage_Ring}. The VCO generates a $50\%$ duty cycle I/Q differential local oscillator.

\begin{figure}[tp]
\centerline{ \includegraphics[width=3.8in]{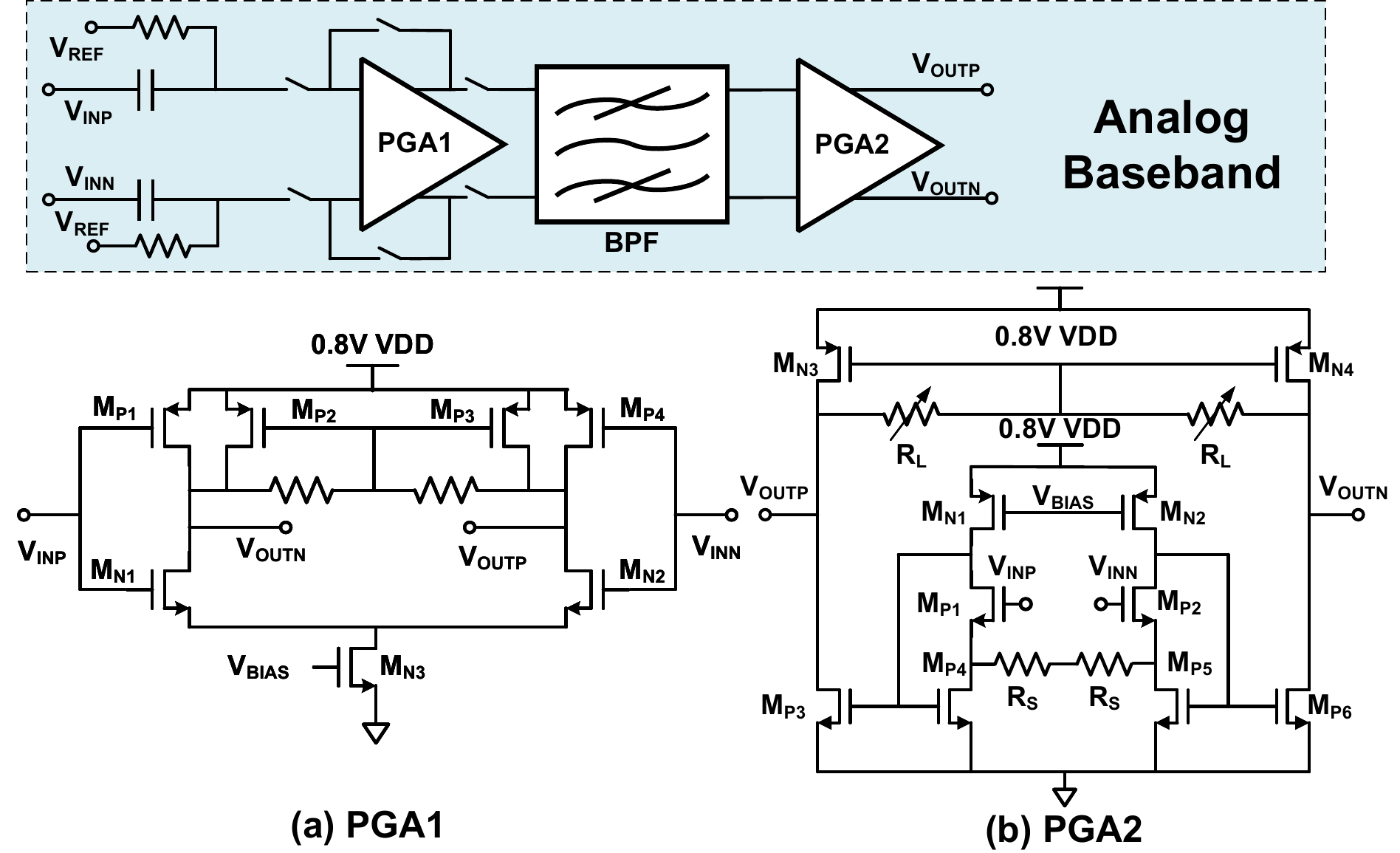} }
\caption{Ultra-low Power Baseband Architecture}
\label{fig9}
\end{figure}

\subsection {Ultra-low Power Baseband}

\begin{figure}[tp]
\centerline{ \includegraphics[width=3.4in]{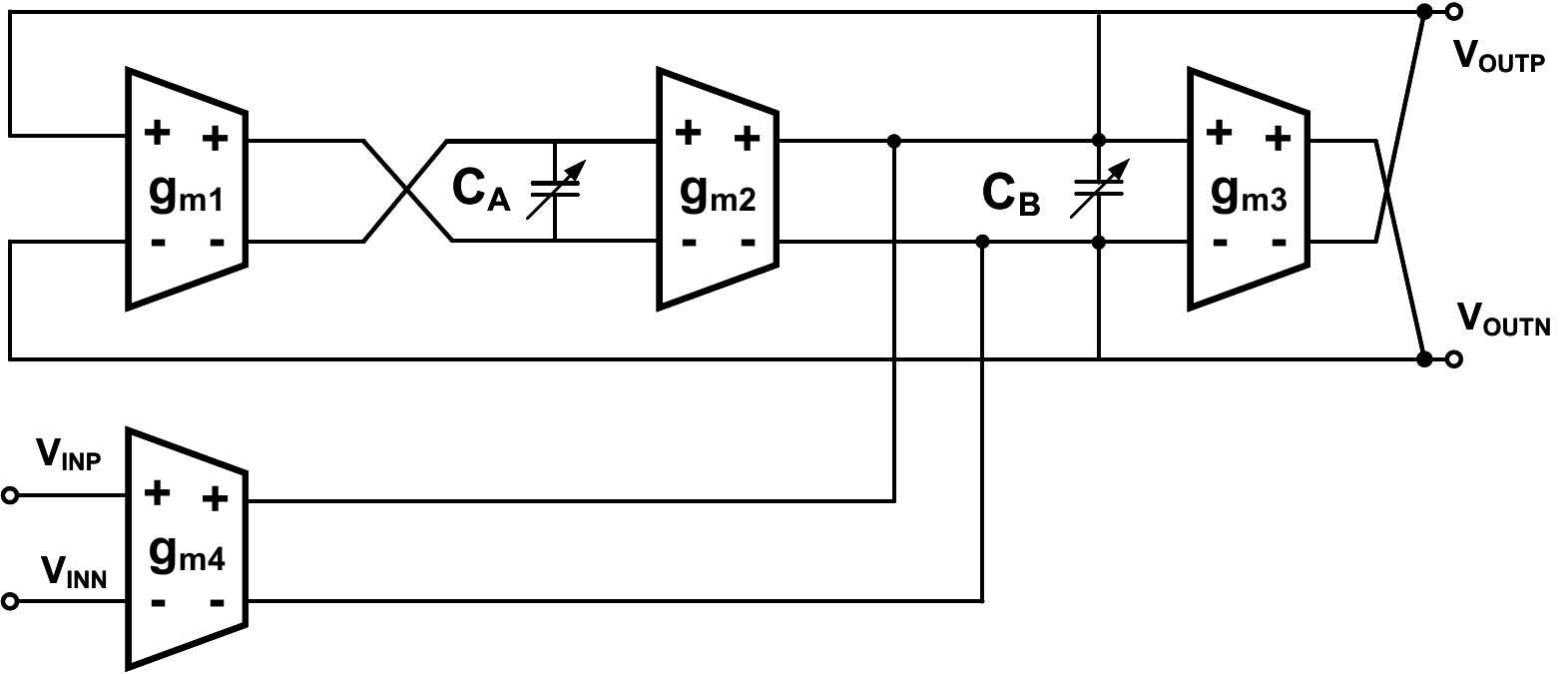} }
\caption{gm-C Bandpass Filter in Baseband}
\label{fig10}
\end{figure}

The low-power analog baseband circuit used in this paper is shown in Fig.~\ref{fig9}. The programmable gain amplifier (PGA) is implemented in two parts: the front PGA (PGA1) and the post PGA (PGA2) based on the connection with the filter. PGA1 provides proper gain to suppress the noise of the subsequent blocks and maintains noise performance. PGA2 can optimise linearity and provide a certain gain to avoid the degradation of the overall receiver linearity due to excessive gain of the front-end blocks. Therefore, the baseband architecture can achieve a good trade-off in terms of noise, linearity, and power. The PGA1 uses an inverter-based current-reuse amplifier structure for low-power, low-voltage design, reducing current consumption and achieving good noise performance. The PGA2 adopts a fully differential trans-conductance enhancement structure and uses local common-mode feedback to reduce power consumption. In addition, the filter circuit uses the gm-C architecture in Fig.~\ref{fig10}.

\section{Measurements and Experimental Results}

\begin{figure}[htp]
  \centerline{ \includegraphics[width=3.4in]{"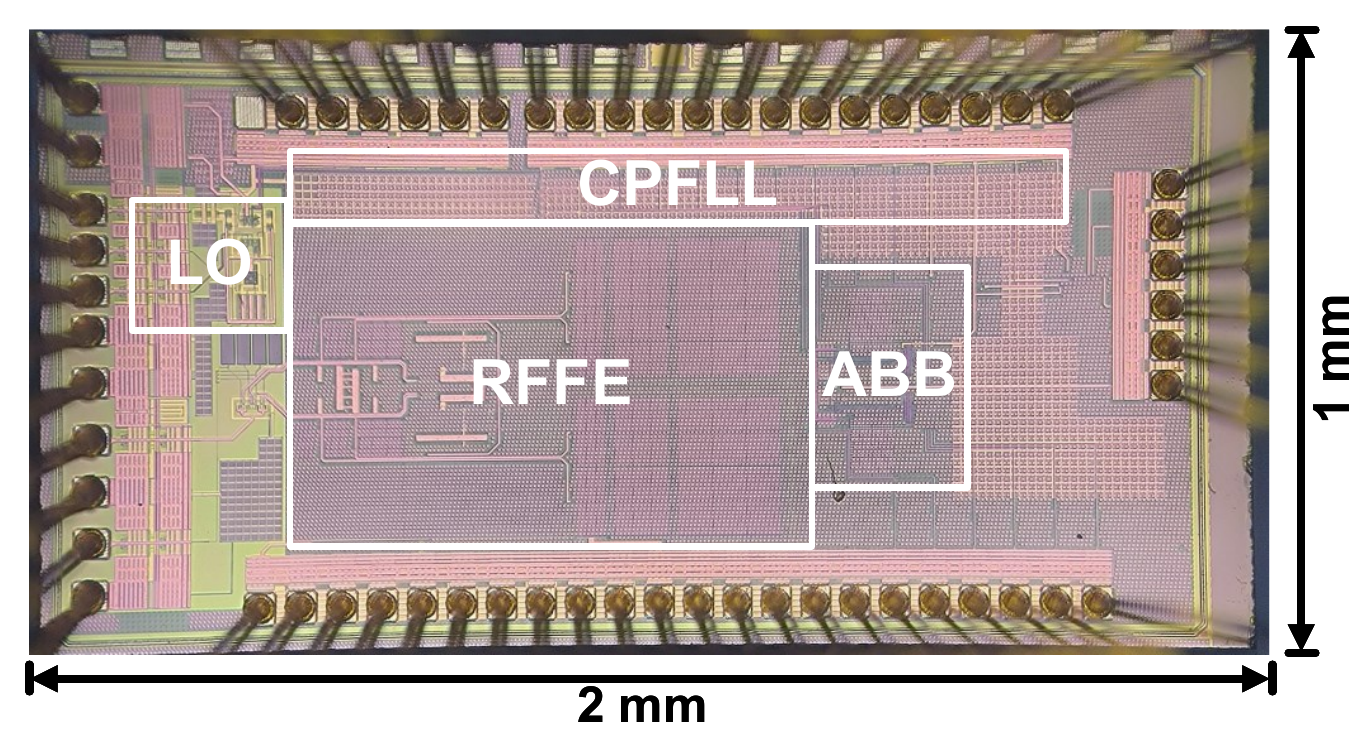"} }
  \caption{The Photos of the Designed A-IoT Prototype Receiver Chip}
  \vspace{-0.2in}
  \label{fig11}
  \end{figure}

As shown in Fig.~\ref{fig11}, the prototype chip is designed and manufactured in the 55 nm CMOS process with a total area of 1 $\times$ 2 $mm^{2}$.

\subsection {S11, Frequency Response and NF of RFFE}

\begin{figure}[tp]
\centerline{ \includegraphics[width=3.2in]{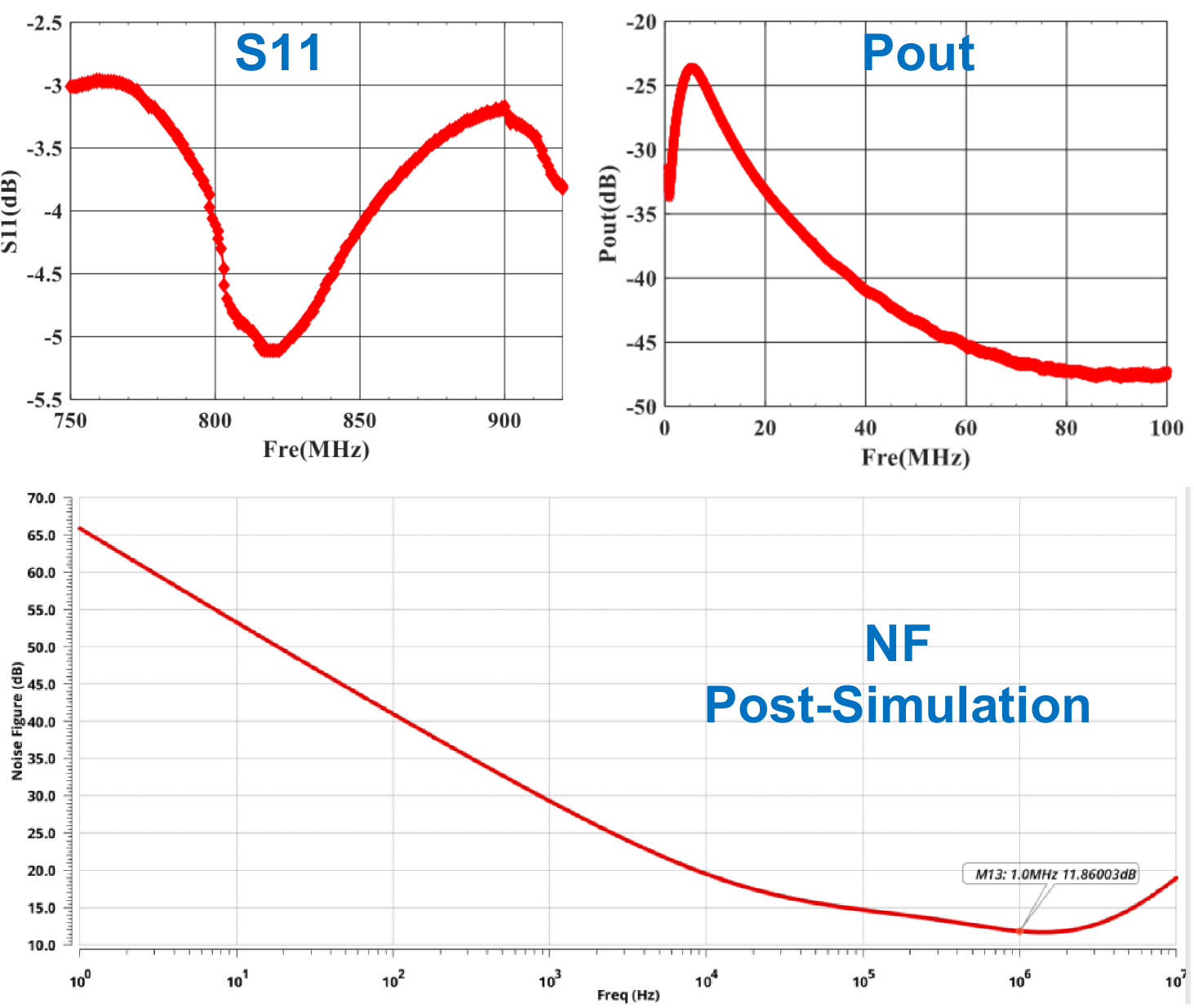} }
\caption{S11, Frequency Response and NF of RFFE}
\label{fig12}
\end{figure}

As shown in Fig.~\ref{fig12}, the S11 shows that the input impedance and the center frequency can be shifted based on the designed Gm-C Gyrator. In addition, changing the local oscillator frequency also changes the center frequency. From the simulation results, the ratio of impedance at IF to the impedance at the image frequency shows an image rejection ratio of 16.7 dB. Based on the frequency response curve, the gain at the out-of-band frequency of 40 MHz is 17 dB lower than 4 MHz, which shows good out-of-band interference suppression. Fig.~\ref{fig12} also shows the noise figure with frequency. With the frequency increasing, the NF decreases due to the flicker noise. At the target 1.035 MHz IF, the NF is about 12 dB.

\subsection {LO Frequency Calibration Loop Dynamic Characteristics}

\begin{figure}[tp]
\centerline{ \includegraphics[width=3.6in]{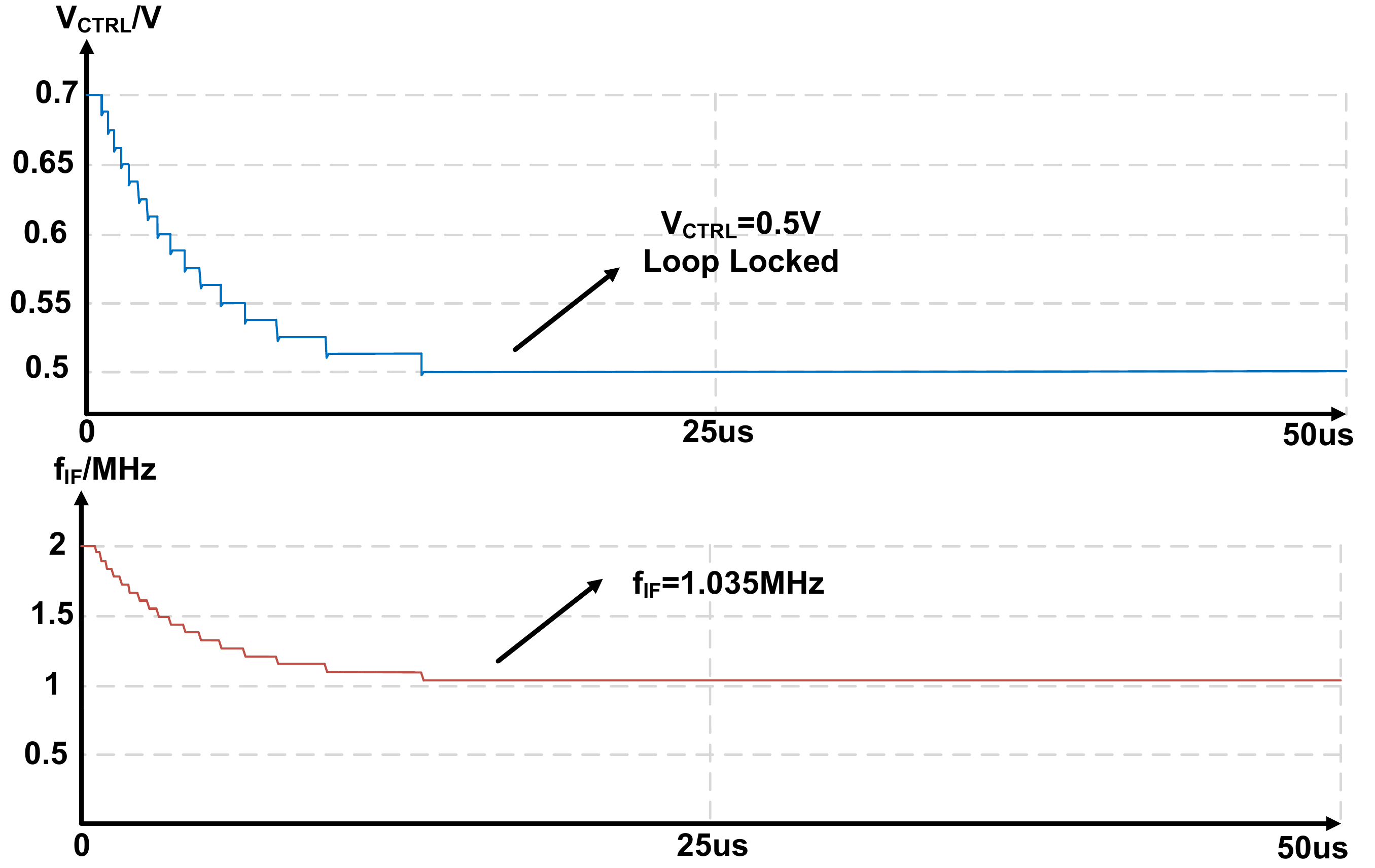} }
\caption{Simulation of Frequency Calibration Loop Dynamic Characteristics}
\label{fig13}
\end{figure}

The time-domain characteristics of the LO frequency calibration loop are evaluated by the behavioural-level simulation model. Fig.~\ref{fig13} shows the $V_{CTRL}$ and $f_{IF}$ curves, which indicate that the loop has typical first-order characteristics. Finally, the loop locks to the target IF with a settling time of about 12 $\mu s$, corresponding to approximately 12 reference cycles. In addition, the loop's operation exhibits discrete-time features due to the RFD and CP.

\subsection {The Receiver Sensitivity Evolution}

The sensitivity of the whole receiver can be estimated according to \eqref{Sensitivity}. Here, it is assumed that the margin is 6 dB, and the required SNR for R2D decoding is 10-15 dB. The evaluation shows that the sensitivity will reach -88 dBm, demonstrating the high sensitivity advantage of narrow-band communication.

\begin{equation}
\setlength{\abovedisplayskip}{0pt}
\setlength{\belowdisplayskip}{0pt}
\label{Sensitivity}
\begin{aligned}
&P_{Sensitivity}\\
&=-174dBm/Hz+10log(BW)+SNR_{min}+NF\\
&+Margin\\
&=-174dBm/Hz+10log(180kHz)+15dB+12dB\\
&+6dB\\
&<-88dBm
\end{aligned}\end{equation}


\section{Discussion and Future Architecture}

\begin{figure}[tp]
\centerline{ \includegraphics[width=3.6in]{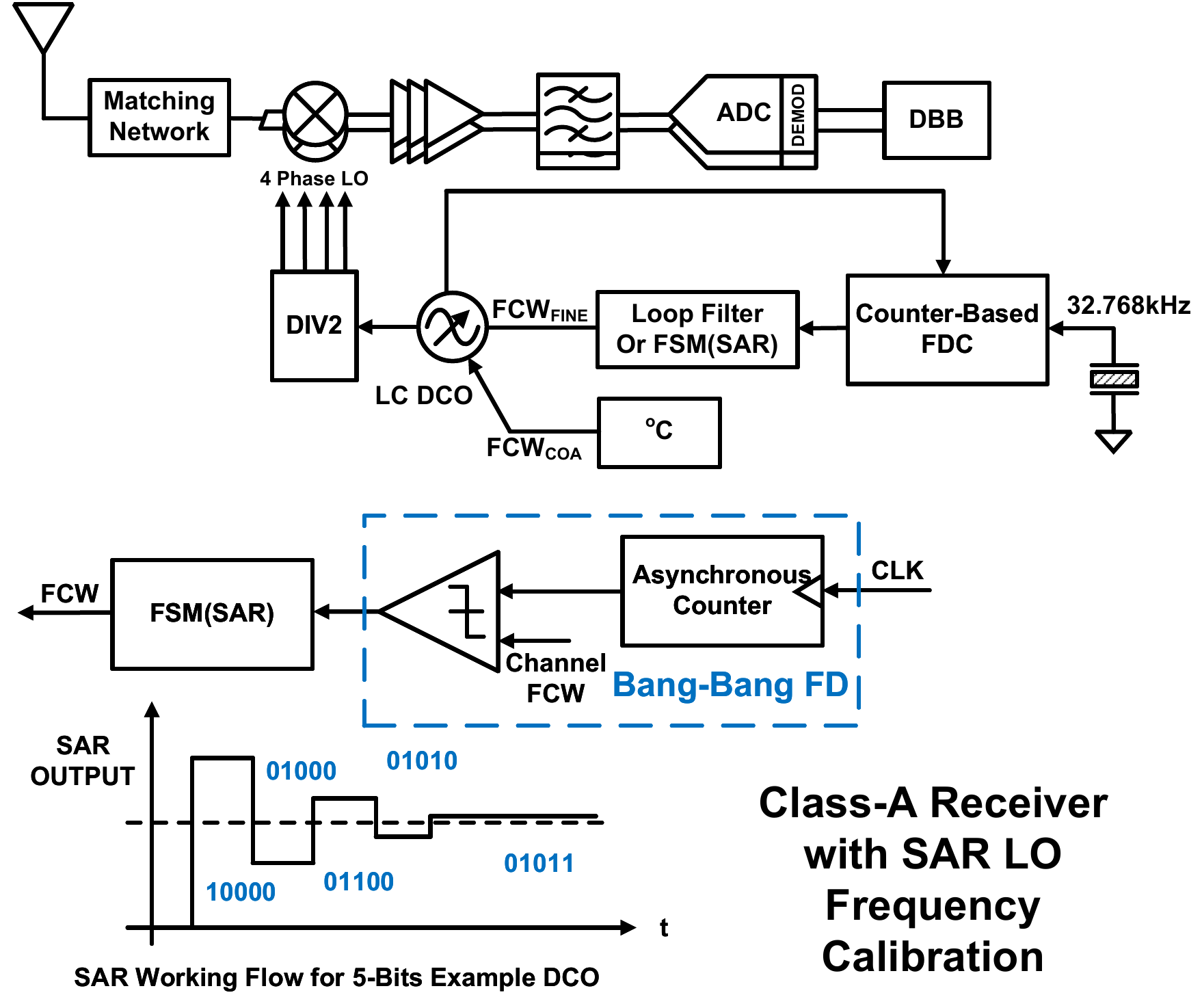} }
\caption{Class-A Receiver with SAR LO Frequency Calibration}
\label{fig14}
\end{figure}

Based on the above discussion, the current analog loop implementation is similar to CPPLL. According to the current trend of radio frequency digitization, the loop clocks can be digitally designed. This paper also shows two feasible digital calibration architectures, the SAR auxiliary LO calibration loop in Fig.~\ref{fig14} and the DFLL based on a low-power asynchronous counter in Fig.~\ref{fig13}.

\begin{figure}[tp]
\centerline{ \includegraphics[width=3.2in]{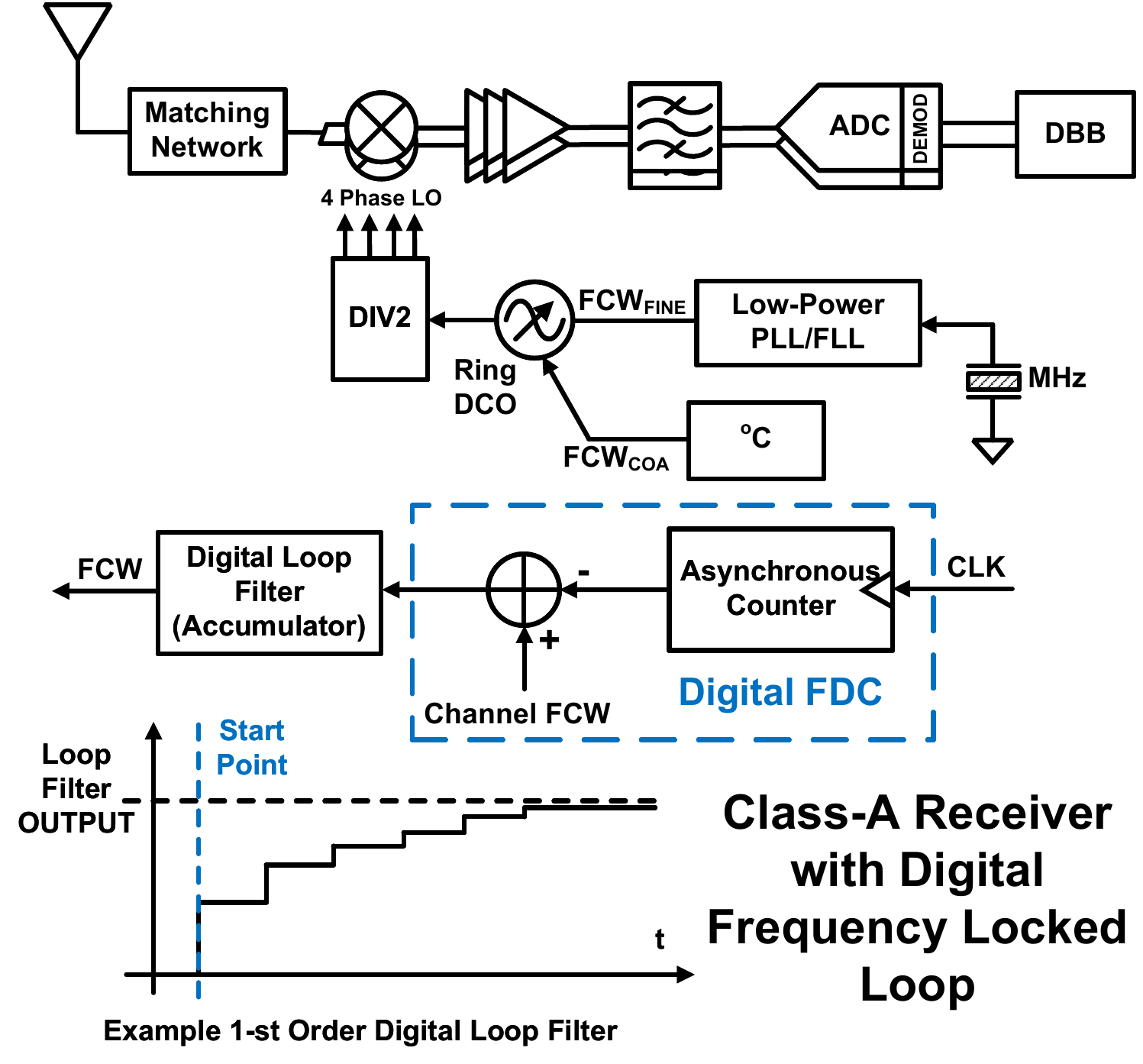} }
\caption{Class-A Receiver with Digital Frequency Locked Loop}
\label{fig15}
\end{figure}

\section{Conclusion}
In conclusion, this paper has comprehensively explored the transceiver design, especially the receiver design for A-IoT. The classification of A-IoT devices into Type-A, Type-B, and Type-C based on power supply and communication activity provides a clear framework for device design. The proposed crystal-less transceiver architecture for Type-B and Type-C devices, featuring an “approximate low-IF” receiver and “carrier-auxiliary IF feedback” LO synthesizer, achieves sub-mW receiver and low-cost clock generation. Experimental results validate the proposed architecture, showing good performance in terms of noise figure and receiver sensitivity. With the trend of radio frequency digitization, digital calibration architectures described in the paper offer potential for further optimization. This paper not only advances the development of “zero power” A-IoT devices but also paves the way for more efficient and widespread IoT connectivity, bridging the digital and physical worlds more seamlessly.

\section*{Acknowledgment}
Copyright \copyright 2024 by [Yuxiao Zhao, Zheng Shen, Shiyu Li, et al.] All rights reserved. This document titled [Considerations on the Design of Transceivers for Ambient Internet of Things] is published as an initial draft version subject to future updates and revisions as the research progresses without prior notice; no alterations amendments or substantial modifications to the Document shall be considered valid unless explicitly approved in writing by all authors and any such changes made without unanimous written consent are strictly prohibited and deemed invalid. The Document lists all authors who have made substantial contributions to the described work including conceptualization design analysis or writing with specific contributions as follows: Yuxiao Zhao is responsible for the design of the RX (Receiver) architecture and associated circuit implementations, Zheng Shen contributed to the design of the RFFE (Radio Frequency Front-End) circuit components, Shiyu Li oversaw the design of the LO (Local Oscillator) circuit subsystem, and the remaining authors provided contributions to the writing editing and critical review supporting the finalization of the content. Users may cite reference or share the Document for non-commercial academic or research purposes provided it is clearly labelled as an "initial draft version" (not a final published work) and all authors with their specific contributions are explicitly credited in derivative uses while commercial use distribution or adaptation without written consent from all authors is strictly prohibited. The date of the initial draft: June 15th, 2024.

%


\ifCLASSOPTIONcaptionsoff
  \newpage
\fi

\bibliographystyle{ieeetr}
\bibliography{ref}

\end{document}